\documentclass[12pt,preprint,superscriptaddress]{revtex4}
\usepackage{psfig}

\begin{document}

\title{Early events in insulin fibrillization 
       studied by time--lapse atomic force microscopy.}

\author{Alessandro Podest\'{a}}
\affiliation{INFM and CIMAINA, Dept. of Physics, University of Milano, via Celoria 16, 20133 Milano, Italy}
\author{Guido Tiana}
\affiliation{Dept. of Physics, University of Milano and INFN, via Celoria 16, 20133 Milano, Italy}
\author{Paolo Milani}
\affiliation{INFM and CIMAINA, Dept. of Physics, University of Milano, via Celoria 16, 20133 Milano, Italy}
\author{Mauro Manno}
\altaffiliation{Corresponding author}
\email{mauro.manno@pa.ibf.cnr.it}
\affiliation{Italian National Research Council, Institute of Biophysics at Palermo, via U. La Malfa 153, 90146 Palermo, Italy }
                                                                                   


\bigskip
\bigskip

{\center \bfseries \Large 
Early events in insulin fibrillization \\ 
studied by time--lapse atomic force microscopy. \\
}

\bigskip
\bigskip

\renewcommand{\thefootnote}{\fnsymbol{footnote}}
Alessandro Podest\'{a} \footnotemark[1], 
Guido Tiana \footnotemark[2],
Paolo Milani \footnotemark[1],
Mauro Manno \footnotemark[3]\footnotemark[4]

\bigskip
\bigskip

\footnotemark[1] 
INFM and CIMAINA, Dept. of Physics, University of Milano, 

via Celoria 16, 20133 Milano, Italy.

\footnotemark[2] 
Dept. of Physics, University of Milano and INFN,

via Celoria 16, 20133 Milano, Italy

\footnotemark[3]
Italian National Research Council, Institute of Biophysics at Palermo,

via U. La Malfa 153, 90146 Palermo, Italy

\footnotemark[4]
To whom correspondence should be addressed. 
E-mail: mauro.manno@pa.ibf.cnr.it

\bigskip
\bigskip
{\bfseries Classification:}

BIOLOGICAL SCIENCES --- Biophysics

\bigskip
{\bfseries Corresponding author:}

Mauro Manno.
Mail address: Italian National Research Council, 

Institute of Biophysics at Palermo,
via U. La Malfa 153, 90146 Palermo, Italy.

Tel.: +39 091 6809305. Fax: +39 091 6809349. 
E-mail: mauro.manno@pa.ibf.cnr.it

\bigskip
{\bfseries Abbreviations:}

AFM, atomic force microscopy.

\bigskip
{\bfseries Manuscript informations:}

Number of text pages: 16  ---  
Number of figures: 3

Number of words in the abstract: 233  ---  
Number of characters in the paper: 40000

\clearpage
{\bfseries
\section*{Abstract}

The importance of understanding the mechanism of protein aggregation 
into insoluble amyloid fibrils relies not only on its medical consequences, 
but also on its more basic properties of self--organization. 
The discovery that a large number of uncorrelated proteins can form, 
under proper conditions, structurally similar fibrils has suggested 
that the underlying mechanism is a general feature of polypeptide chains.
In the present work, we address the early events preceeding 
amyloid fibril formation in solutions of zinc--free human insulin
incubated at low pH and high temperature.
Aside from being a easy--to--handle model for protein fibrillation, 
subcutaneous aggregation of insulin after injection is a nuisance 
which affects patients with diabetes. 
Here, we show by time--lapse atomic force microscopy (AFM) 
that a steady-state distribution 
of protein oligomers with an exponential tail is reached within few
minutes after heating. This metastable phase lasts for few hours
until aggregation into fibrils suddenly occurs.
A theoretical explanation of the oligomer pre--fibrillar distribution
is given in terms of
a simple coagulation--evaporation kinetic model, in which 
concentration plays the role of a critical parameter.
Due to high resolution and sensitivity of AFM technique, 
the observation of a long-lasting latency time should be 
considered an actual feature of the aggregation process, and not
simply ascribed to instrumental inefficency.
These experimental facts, along with the kinetic model used,
claim for a critical role of thermal concentration fluctuations
in the process of fibril nucleation.
}

\clearpage
\section*{Introduction}

Self--assembly of proteins or peptides into linear elongated structures
known as amyloid fibrils is a conserved feature accompaning the clinical 
manifestation of many pathologies, 
such as Systemic Amyloidosis or several neurodegenerative diseases
(Alzheimer's disease, Transmissible Spongyform Encephalopathy, etc.)
\cite{Kelly1996}.
In several cases, fibril formation is regarded as the onset and the cause
of such diseases  (``Amyloid Hypothesis'') \cite{Hardy2002}.
More in general, a large number of uncorrelated proteins 
share the possibility
to assemble into similar fibrillar structures under appropriate conditions,
that typically favour non native conformations
\cite{Chiti2002,Stefani2003}.
Therefore, the study of fibrillation kinetics
is important in order to understand
the processes and the interactions involved in 
amyloid self-assembly and
to design molecular inhibitors.

The 51--residue hormone insulin has long been known 
to form fibrils if heated at low pH \cite{Langmuir1940,Waugh1941},
that is when monomeric or dimeric forms are promoted \cite{Doty1953,Brange1997}.
Indeed, insulin is protected from fibrillation by assembling into Zn--hexamers 
during {\it in vivo} storage or in artificially delivery systems 
\cite{Brange1987,Lougheed1980}.
In acidic condition, insulin aggregation proceeds mainly via three steps
\cite{Waugh1953,Waugh1954,Waugh1957}: formation of active centers 
(nucleation), elongation of these centers to fibrils (growth), and floccule formation 
\cite{Krebs2004}.
This is a typical scheme for protein polymerization \cite{Oosawa1975}
or amyloid formation \cite{Harper1997,Lomakin1996,Lomakin1997}.
More recenty, the structure of insulin fibril has been shown to resemble
that of typical amyloid fibrils with the characteristic cross-$\beta$ structure
\cite{Nettleton2000,Bouchard2000,Tito2000,Jimenez2002}.

In order to understand the molecular mechanism which is responsible 
for the uprise of fibrils, it is necessary to get insight into 
the early stages of the process. 
Observation of partially folded intermediate conformations in conditions preceeding
insulin fibril formation provided a molecular insight of the interactions involved
\cite{Hua1991,Brange1997b,Whittingham2002,Ahmad2003,Ahmad2004,Hua2004},
yet the onset of aggregation and the causes leading
to fibril nucleation and elongation are not clearly understood.

The early stages of fibrillogenesis are, in general, difficult to investigate, 
due to the inherent instability of such systems. 
Quenching the incubating solution to low temperature allows 
to perform molecular weight filtering and circular dichroism experiments 
\cite{Jarvet2000}, but the information one obtains concerns conditions 
different from the incubating ones. 
Light scattering \cite{Lomakin1996,Berne1976} is particularly suited to
detect either large supramolecular aggregates 
or protein size objects at sufficiently high mass concentration,
and consequently it misses the early events in fibrillation kinetics. 
Neutron scattering has been used to detect small fibrillar precursors
\cite{Yong2002}, but needs long measurements, and thus one can obtain 
but time--averaged quantities.
Atomic force microscopy (AFM) is a technique able to detect fine--grained 
features of samples deposited on a substrate (the resolution corresponding 
to the inverse curvature of the tip, that is $\sim 5$ nm). 
Time-lapse AFM have been extensively used to observe 
the stucture and growth of amyloid fibrils.

In the present work, we performed AFM experiments
during fibrillation of human insulin.
In particular, we focus on the early stages preceeding the 
observation of mature fibrils. 
In order to explore with sufficient time-resolution the lag phase,
we used zinc--free recombinant human insulin, 
since in this case fibril formation takes place 
on the time scale of hours \cite{Brange1997,Hua2004},
and it is slower than that of the best studied bovine insulin
\cite{Nielsen2001,Nielsen2001b,Nielsen2001c,
Dzwolak2003,Jansen2004,Jansen2005}.
AFM snapshots at different times show a distribution of ellipsoidal
oligomeric aggregates, consistent with analogous finding 
in other amyloidogenic systems, as 
the Alzheimer's amyloid-$\beta$(1-40) peptide
\cite{Walsh1997,Huang2000,Westlind2001,Hoshi2003}
or other proteins \cite{Relini2004,Xu2001}.
After 4 hours of incubation, ellipsoidal protein oligomers
disappear from AFM images and amyloid fibrils of different length are detected,
with a structure analogous to that observed for bovin insulin fibers
\cite{Jimenez2002,Jansen2005}.
Such abrupt change in aggregate distribution and shape occurs within the 
experiment time resolution that is 30 minutes.

A main result obtained from our experiments is that the 
oligomer distribution is stationary during the lag-phase
and it exhibits an exponential tail.
The median values of this distribution are consistent with 
electro-spray mass-spectrometry experiments performed on 
bovin insulin in analogous conditions \cite{Nettleton2000},
but also larger oligomer, up to several tens, are involved.
This metastable phase can be explained by a 
coagulation--evaporation process that has been proposed for
colloidal aggregation \cite{Krapivsky1996}.
As to this model, the existence of a stationary 
oligomer distribution 
is critically controlled by protein concentration. 
Consequently, small local concentration fluctuations are enough 
to make the system cross to the 
dynamical phase characterized by large ``elongated'' growing clusters.

Our work is thus in harmony with experimental observations
\cite{Piazza2002,Stradner2004} and theoretical studies \cite{Sciortino2004,Mossa2005}
of protein clusters distributions in conditions promoting protein crystallization.
Now, the present results shed a new light into the current view of fibril
nucleation, assigning a relevant role to thermal fluctuations 
and to protein-protein interactions leading to cluster formation
rather than to physical fibrillar precursors.

\section*{Materials and Methods}

{\bfseries Sample preparation.} 
Recombinant human insulin powder (purchased from Sigma Chemical Co. 
and used without further purification)
was directly dissolved at 5 $^\circ$C in buffer solution 
(50 mM KCl/HCl in Millipore SuperQ water, pH 1.6 at 60 $^\circ$C).
The protein solution was gently stirred, filtered through 0.22 mm 
Millex-GV (Millipore) filter into glass cells, and incubated 
at 60 $^\circ$C.
Insulin concentration was 200 $\mu$M as measured by UV absorption 
at 276 nm using an extinction coefficient of 1.0675 for 1.0 mg/ml. 
The final concentrations were consistent with those
calculated by weigthing insulin powder, thus confirming 
that essentially no material was lost through filtering
and that insulin was efficiently dissolved.
After given time intervals 10 $\mu$l of incubated protein solution were 
diluted into 1 ml buffer solution, quenched to 0 $^\circ$C to
rapidly inhibit further aggregation, and used for
atomic force microscopy experiments.
All chemicals were analytical grade.

{\bfseries Atomic force microscopy (AFM).} 
A few $\mu$l of the insulin solution were dropped
onto a freshly cleaved mica substrate (quality ruby muscovite). 
After few minutes, the sample was washed dropwise with Millipore SuperQ water, 
and then dried with a gentle stream of dry nitrogen. 
Images of the protein aggregates were recorded with a 
Multimode Nanoscope IIIa AFM (Veeco Instruments, Santa Barbara, CA, USA), 
operating in Tapping Mode inside a sealed box where 
a dry nitrogen atmosphere was maintained. 
We used rigid cantilevers with resonance frequencies of about 300 kHz, 
and equipped with single crystal silicon tips with nominal radius of
curvature 5-10 nm. 
Typical scan size was 500x500 nm$^2$ (512x512 points), and scan rate 1-2 Hz.

{\bfseries Static Light Scattering.}
Immediately after preparation, samples were placed 
in a thermostated cell compartment of
a Brookhaven Instruments BI200-SM goniometer,
equipped with a 100 mW Ar laser tuned at $\lambda_0=514.5$ nm.
The temperature was set at 60 $^\circ$C and controlled within 0.05 $^{\circ}$C
with a thermostated recirculated bath.
Scattered light intensity at 90$^{\circ}$ 
was measured by using a Brookhaven BI-9000 correlator.
Absolute values for scattered intensity (Rayleigh ratio)
have been obtained by normalization
with respect to Toluene, whose Rayleigh ratio at 514.5 nm
was taken as $32 \cdot 10^{-6} cm^{-1}$.

\section*{Results}

\begin{figure}[h]
\centerline{\psfig{file=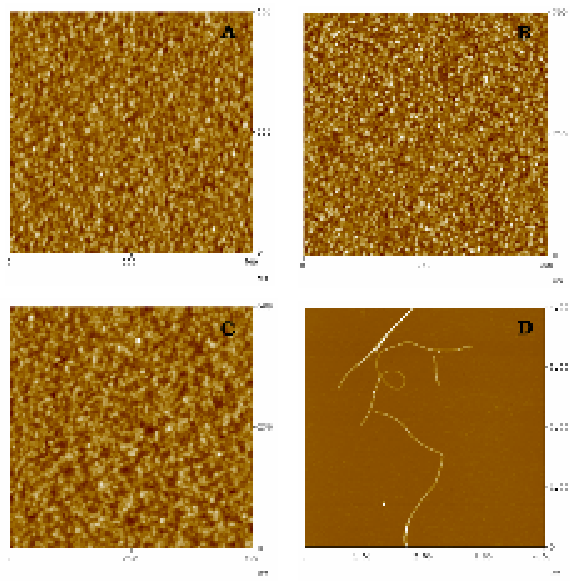,width=8.6cm}}
\caption{
         Snapshots of insulin aggregation kinetics at 60 $^{\circ}$C 
         monitored by kinetic AFM. Times elapsed after incubation:  
         (A) 1 min. (B) 180 min. (C) 250 min. (D) 540 min.
         The vertical color scale is (A)-(C) 5 nm, and (D) 30 nm
        }
\label{fig1}
\end{figure}

{\bfseries Time--resolved AFM.}
Our procedure to investigate early stages of insulin fibrillation
consists in incubating the protein in a test tube, 
extracting samples every 30 minutes, 
depositing on a substrate and scanning it with the AFM 
(which takes a time of the order of minutes). 
Thus, we obtain snapshots of the aggregation intermediates until 
fibrils are formed.
Several AFM images of each sample, representative of a given incubation time, 
were recorded. 
This allowed collecting the topographic data of about 10$^4$ aggregates 
for each incubation time. 
Snapshots of the system from the beginning of the incubation 
(defined as time zero) up to nine hours are displayed in Fig. \ref{fig1}. 
Such snapshots indicate that there are oligomers, 
but not fibril--like structures (cf. Fig. \ref{fig1}A--C), 
in the first four hours, until fibrils suddenly appear at time 280 min.  
(cf. Fig. \ref{fig1}D). 
The overall process can thus be divided into a long metastable phase, 
a nucleation event and the growth of the fibrils. 
Note that the growth phase is much faster than the metastable phase, 
the fibrils having incorporated all oligomers within the time resolution 
of the experiment, that is 30 minutes.

{\bfseries AFM data analysis in the early stages of kinetics.}
An home made software was used for
detecting the edges of the protein aggregates in the AFM maps 
[\textit{M. Marino, A. Podest\'a, P. Piseri et al..., unpublished}]. 
The binary maps obtained were then processed using the Image Processing 
Toolbox of Matlab (The Mathworks, Inc.) and the average distributions 
of aggregate areas were obtained, as shown in Fig. \ref{fig2}A--C. 

Deconvolution of the tip shape from AFM images is a critical issue
in any quantitative study of biological samples.
Deconvolution algorithms are likely to introduce artefacts in the data,
expecially when the basic features in the AFM maps
are nanometer sized. Moreover, the morphology of our
system, a quasi two-dimensional close arrangement of nanometer sized objects,
without gaps in between, does not permit to apply simple
deconvolution formula to the distribution of areas \cite{Odin1994}.
These formula apply to the case of parabolic-spherical tips
scanning isolated objects lying on a flat reference plane.

We have thus processed raw AFM images without applying any deconvolution. 
We expect indeed reduced convolution effects, because
the tip does not penetrate deeply down to the substrate, 
but only sense the outmost surface of the protein layer. 
This insures only negligible lateral contact of the tip and consequently 
reduced loss of resolution. In addition, the underestimation 
of the area of the aggregates caused by the erosion of binary maps 
operated by the edge detection algorithm tends to compensate the opposite 
effect produced by the tip shape convolution.

To show that the effects of tip convolution are  negligible, 
we analyzed several AFM images of highly diluted samples,
where isolated aggregates lying on the flat mica surface are visible 
(about 20 complexes every 500x500 nm$^2$). 
These model samples were pre--processed using standard deconvolution algorithms;
we used the formula $w'=w-2\sqrt(2hR_{tip})$ - parabolic tip on a step - 
where $R_{tip}$ is the tip radius (assumed $R_{tip}$~3 nm), 
$h$ is the step height ($h$~1.1 nm, the average aggregate height 
extracted by the AFM images), $w$ and $w'$ are the apparent and deconvoluted 
widths of the observed features \cite{Odin1994}.
The resulting distribution of areas were in good agreement with those 
obtained from the non-deconvoluted AFM images (data not shown). 
In particular, the median and standard deviation of the areas were 
27 $\pm$ 35 nm$^2$ accordingly, to be compared with the average values of 
30 $\pm$ 22 nm$^2$, extracted from the raw AFM images of
concentrated samples.

Quantitative estimation  of aggregate size from areas rather than 
from heights is more reliable because the peculiar vertical interaction 
of the AFM tip with biological samples usually leads to underestimation 
of the true height. The same effect is caused by the close packing of 
insuline aggregates in relatively concentrated samples, which keeps the tip 
from getting in touch with the flat reference substrate. 
Processing AFM images of concentrated samples, however,  allowed 
collecting a large statistics,
required to have a stable fit of the area distributions.

{\bfseries Shape of oligomers.}
The shape of the aggregates can be characterized by mean of their eccentricity.
The eccentricity of the protein aggregates was also evaluated 
from the binary maps using the same Matlab toolbox. 
Eccentricity is defined as $\sqrt{1-(a/b)^2}$, 
$a$ and $b$ being the minor and major axis, accordingly.
This parameter is expected to be 0 for a circle, and 1 for a segment.
Correlations of eccentricity and areas are shown in Fig. \ref{fig3},
which show that also this feature of the system 
is stationary in the metastable phase. 
Aggregates have a mean eccentricity of 0.75, that stands for a ratio between 
large and small axis of about 1.5.
Larger size aggregates have a larger eccentricity than smaller aggregates,
thus evidencing a preferencial unidimensional (fibrillar) growth 
for clustering proteins,
consistent with recent theoretical findings on colloid clusters with both short range
attraction and long range repulsion \cite{Sciortino2004,Mossa2005}.

Because of the rounding effect of tip convolution, the measured eccentricity is, 
at most, an underestimate of the actual one.

{\bfseries Average mass of insulin oligomers at the onset of kinetics.}
Light scattering experiments were performed 
immediately after incubation at 60 $^\circ$C.
Measurement of the intensity scattered
at 90$^{\circ}$ (scattering vector $q=23 \mu m^{-1}$) provides
the Rayleigh ratio $I_R(q)$ that is related to the weight average
molecular mass $M_w$ by the relation:
$I_R(q)=4\pi^2\tilde{n}^2(d\tilde{n}/dc)^2\lambda_0^{-4}N_A^{-1}cM_wP_z(q)$,
with $c$ mass concentration, $\tilde{n}$ medium refractive index,
$\lambda_0$ incident wavelength, $N_A$ Avogadro's number, and
$P_z(q)$ z-averaged form factor \cite{Berne1976}.
By taking $(d\tilde{n}/dc)=0.18~cm^3 g^{-1}$, and $P_z(q)=1$
(since the initial size of solutes is much smaller than $q^{-1}$),
we obtain an average molecular mass of 23$\pm$5 kDa.
Considering that the molecular mass of a single insulin molecule
is 5806 Da, the soluble oligomers found at the onset of kinetics
are made up of about 4$\pm$1 insulin molecules. 
Note, however, that the mean aggregation
number obtained by light scattering measurements corresponds to the ratio
between the second and the first moment of oligomer distribution \cite{Berne1976}, 
and it gives no information on the actual distribution shape.

{\bfseries Oligomer distribution preceeding amyloid formation.}
Volumes of imaged objects were derived
from calculated areas and eccentricities, 
under the assumption that the aggregates are prolate ellipsoids.
Aggregation numbers $n$ are obtained 
by using the relation $V=V_0n^{1/d}$, where 
$V_0=14.1 nm^{3}$ is the van der Waals volume of an insulin monomer, 
including a layer of water, derived from the x-ray structure \cite{Baker1988},
and $d$=2.68 is an effective fractal dimension that accounts for the scaling 
between mass and size of aggregates. 
The value $d$=2.68 is derived from 
x-rays and light scattering data 
\cite{Baker1988,Kadima1993,Pedersen1994}
on oligomers of zinc--free insulin at high pH. 
Note that zinc-free insulin is not tightly packed 
nor it is assembled into toroidal shaped hexamers as zinc insulin. 
We have checked that 
by assuming an effective fractal dimensions between 2 (enough loose aggregates)
 and 3 (space filling objects), 
the shape of oligomer distribution is not significantly altered,
that is the distribution shape is robust
with respect to different reasonables choice of molecular packing.
This distribution implies that aggregates built out of up to 50 monomers 
are detectable in the initial stages of aggregation.
The large size of these oligomers is in agreement with the 
micellar precursors identified in ref. \cite{Lomakin1996}
in the case of Alzheimer's amyloid--$\beta$ peptide.

The distribution of oligomer aggregation numbers at different times 
in the metastable phase is displayed in Fig. \ref{fig2}D--F. 
All the curves well overlap indicating that the distribution of size is stationary. 
The median aggregation numbers $n_m$ are 5.9, 4.9, and 6.7
respectively for the three cases shown in the figure.
The tail of such distributions can be fit by an exponential 
of the kind $\exp(-n/n_m)$ (cf. Fig. \ref{fig2}), 
where $n_m$ is the median aggregation number. 

\section*{Discussion}

\begin{figure}[h]
\centerline{\psfig{file=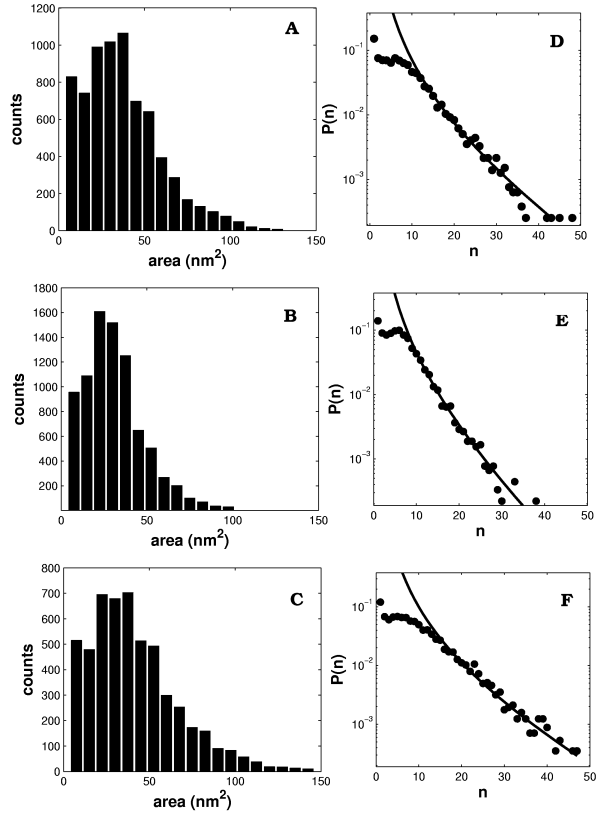,width=8.0cm}}
\caption{
        Oligomer distributions in the course of kinetics.
        (A)--(C) 
        Counts of areas observed in AFM images 
        of figures \ref{fig1}A--C respectively. 
        (D)--(F) 
        Frequency of occurrence of aggregation numbers of objects 
        observed in AFM images of figures \ref{fig1}A--C respectively. 
        Solid lines are fit by expression \ref{eq:redner}
        }
\label{fig2}
\end{figure}

\begin{figure}[h]
\centerline{\psfig{file=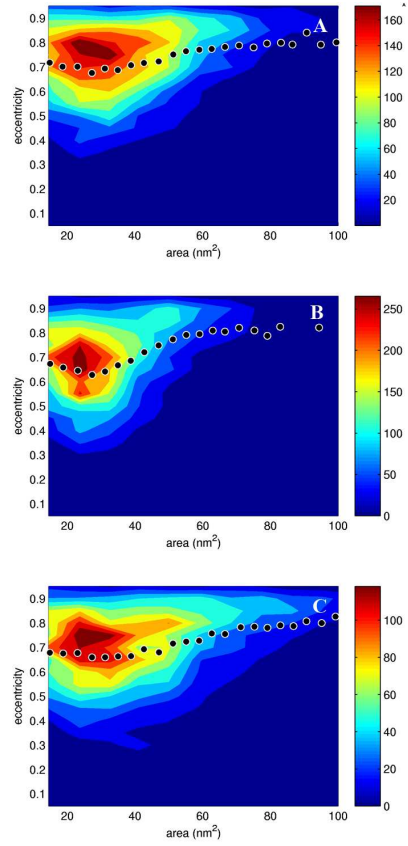,width=5.3cm}}
\caption{
        Size --- eccentricity correlation.
        (A)--(C) 
        Correlation of eccentricity and areas of objects observed in AFM images 
        of figures \ref{fig1}A--C respectively. 
        Dotted curves represent average eccentricity versus aggregate area
        }
\label{fig3}
\end{figure}

{\bfseries Kinetic model for oligomer distribution.}
The most evident feature of the distributions of oligomer size 
and aggregation number
shown in Fig. \ref{fig2} is that they reach a steady state within the time 
detectable from the experiment (i.e. few minutes). 
A steady state means that, unlike diffusion--limited or reaction-limited mechanisms 
which regulate the assembly of larger aggregates \cite{Lomakin1997}, 
in the present case we deal with an ``evaporation''  process 
(i.e., monomers leaving the aggregates) which competes with ``coagulation''.

A mechanism for protein association which account for 
both aggregation and evaporation processes can be outlined in
the framework of classical coagulation theory \cite{Chandrasekhar1943}. 
If we call $\rho_n(t)$ the number concentration of aggregates built out of 
$n$ monomers at time $t$, the rate equation of the system reads:
\begin{eqnarray}
\label{eq:smoluchowski}
\nonumber
\dot{\rho}_n(t)=
\frac{1}{2}\sum_{i+j=n}K_{ij}\rho_i(t)\rho_j(t)-\rho_n(t)\sum_jK_{nj}\rho_j(t) \\
+\lambda_{n+1}\rho_{n+1}(t)-\lambda_{n}\rho_n(t)+\delta_{n,1}\sum_j\lambda_{j}\rho_j(t)
\end{eqnarray}
where dotted quantities refer to time derivatives.
The first two terms in the right hand side of the latter equation are
respectively the production and loss of $n$-mers by coagulation
of two clusters of $i$ and $j$ proteins, 
while the other terms describe the ``evaporation'' of one monomer from 
a cluster of $n+1$ proteins into a cluster of $n$ proteins and a single protein. 
Here, we are including no nucleation term, and we are also 
assuming that three-body effects can be neglected.

The simplest solution of such equations has been provided by 
Krapivsky and Redner \cite{Krapivsky1996} by taking mass independent rate costants,
$K_{ij}=K$ and $\lambda_{i}=\lambda$, 
and assuming that only monomers are present at time zero, that is
$\rho_n(0)=cM_0^{-1}\delta_{n,1}$, where $c$ is the total
mass concentration and $M_0$ is the mass of a monomer.

The model displays two behaviours, controlled by 
the parameter $\mu=K\lambda^{-1} cM_0^{-1}$, that is by
the ratio between the coagulation and the evaporation rate constants
and by the initial concentration of monomers.
At low protein concentration ($\mu<1$) 
the system displays a steady state distribution $P(n)=\rho_n /\Sigma \rho_n$ 
with an asymptotic exponential tail:
\begin{equation}
\label{eq:redner}
P(n)=x^{n-1}
\frac{\Gamma\left(n-\frac{1}{2}\right)}{\Gamma\left(n+1\right) \Gamma\left(\frac{1}{2}\right)}
\left[ 1-\frac{n-\frac{1}{2}}{n+1} x \right]
\end{equation}
where $x=\mu(2-\mu)$.
At $\mu=1$, the system display a power-law distribution, 
while at higher concentrations ($\mu>1$) it does not display any steady state, 
the typical cluster growing linearly in time.

The tails of the distributions shown in Fig. \ref{fig2}D--F 
are well fit by equation \ref{eq:redner}, indicating that the system is 
in the low--concentration regime. 
For the three ditributions
one obtains respectively $\mu$=~0.71,~0.66,~0.73. 
The mass averaged mean aggregation number $n_z$, which is
accessible through scattering experiments and is found to be $4\pm 1$,
can be expressed in terms of the present model as the ratio between the
second and the first moment of the distribution: 
$n_z=1/(1-\mu)=3.3 \pm 0.4$.

Due to the large value observed for the
parameter $\mu$, one could speculate that local fluctuations in the
density of monomers could be the triggering mechanism behind the onset
of fibril formation, akin to what proposed for crystal nucleation
\cite{tenWolde1997,Manno2004b}. 
Note that this does not imply a symmetry breaking,
since the metastable aggregates already display a pronounced
eccentricity.

The kinetic model used need no assumption concerning 
thermodynamic equilibrium. Notwithstanding, it is 
interesting to consider the free-energy change involved 
in the clustering process if one assumes a ``metastable''
equilibrium condition. In particular, 
one can define the free energy $\Delta G_{n}$ associated with
the addition of one monomer to a cluster of n proteins, as:
$\Delta G_{n}=-k_BT \mathrm{ln}(f_{n+1}/f_nf_1)$, where $k_B$ is 
the Boltzmann constant and $f_n$ is the activity of a cluster of $n$ proteins.
If we take the activity as $f_n=c_n/c$, with $c_n$ mass concnetration of the 
n-mers and $c$ total concentration, we obtain for an infinitely large cluster: 
\begin{equation}
\label{eq:g_infty}
\frac{\Delta G_{\infty}}{k_BT} = -\mathrm{ln}\frac{2\mu}{1-\frac{1}{4}\mu(2-\mu)} 
\end{equation}

>From our analysis we obtained  $\Delta G_{\infty}=-0.6k_BT$.
Therefore, the free-energy related to the growth of a large cluster or fiber is
easely accessible through a thermal fluctuations.

This gives a rationale for the fact that in insulin as well as in other protein solution
a change in temperature or in solvent conditions can trigger fibril formation
\cite{Manno2004}.

{\bfseries Conclusive remarks.}
In the present work, the early stages of human insulin fibrillation have been 
monitored by time--lapse AFM, a techniques with high resolution and sensitivity.
Experimental observations and theoretical modeling highlight an interesting
scenario of the nucleation mechanism preceeding amyloid fibrillation.
i) Experiments show that a steady--state distribution of protein oligomers 
with an exponential tail is present in solution up to the abrupt formation
of amyloid fibrils (Fig. \ref{fig1} and \ref{fig2}).
ii) Oligomer distribution can be explained by a kinetic model 
that combines coagulation and evaporation events (Fig. \ref{fig2}D-F). 
As to this model, the formation of ``non-stationary'', growing aggregates 
is controlled by monomer concentration. 
In the present case, concentration is below the critical value,
yet sufficiently high to allow ``above--threshold'' thermal concentration fluctuations.
iii) Pre-fibrillar oligomers exhibit a marked eccentricity (Fig. \ref{fig3}),
denoting that the symmetry-breaking implied by the existence of fibrillar aggregates
is already occurred before fibrillation.  
Indeed, it is reasonably related to a ``fast'' conformational change
\cite{Brange1997b,Tito2000,Hua2004}.
The existence of prefibrillar precursor acting as aggregation nuclei 
has been widely observed in amyloid formation 
both as pre-existing seeds and as actual self-assembled nuclei
\cite{Harper1997}.
The present results point out that along with the existence of such precursors
local density fluctuations may play a critical role in the nucleation mechanism
and trigger amyloid fibrillogenesis.

\section*{Aknowledgements}

We gratefully aknowledge the help of L. Finzi.
We thank E. Craparo, P.L. San Biagio, C. Rishel and F. Librizzi
for collaboration and access to unpublished data. 
One of the author (M.M.) thanks V. Martorana for countless discussions.
This work was partially supported by the Italian
Ministero della Salute through the projects
{\em Neuropatie animali:
analisi molecolari e funzionali della proteina prionica
in razze bovine siciliane}
and
{\em Deposito della beta amiloide nella membrana cellulare:
ruolo degli ioni metallici e dei radicali liberi}.


\clearpage
\section*{Figure captions}

\end{document}